\documentclass[letterpaper,twocolumn,10pt,review,anonymous]{article}
\usepackage{usenix,epsfig,endnotes}
\usepackage{upgreek}
\usepackage{pifont}
\usepackage{amsmath}
\usepackage{multirow}
\usepackage{multicol}
\usepackage{makecell}
\usepackage{booktabs}
\usepackage{xcolor}
\usepackage{subcaption}
\usepackage{array}
\usepackage{algorithm2e}
\usepackage{amsmath}

\usepackage{booktabs}

\newcommand{\etal}{{\em et al.}}

\newcommand{\sysname}[0]{{RobustRL}}
\newcommand{\corrauth}{$^\dagger$}

\begin{document}

\date{}

\title{\Large \bf Role-Based Fault Tolerance System for LLM RL Post-Training}

\author{
  Zhenqian Chen\textsuperscript{1}, 
  Baoquan Zhong\textsuperscript{2}, 
  Xiang Li\textsuperscript{2}, 
  Qing Dai\textsuperscript{1}, 
  Xinkui Zhao\textsuperscript{1}\corrauth,\\
  Miao Ye\textsuperscript{1}, 
  Ren Cheng\textsuperscript{2}, 
  Lufei Zhang\textsuperscript{3}, 
  Jianwei Yin\textsuperscript{1}\\
  \textsuperscript{1}Zhejiang University
  \textsuperscript{2}Unaffiliated\\
  \textsuperscript{3}State Key Laboratory of Mathematical Engineering and Advanced Computing, China
}



\maketitle

\thispagestyle{empty}






\renewcommand{\thefootnote}{} 
\footnote{\corrauth Corresponding author.}
\renewcommand{\thefootnote}{\arabic{footnote}} 



\begin{abstract}

RL post-training for LLMs has been widely scaled to enhance reasoning and tool-using capabilities. However, RL post-training interleaves training and inference workloads, exposing the system to faults from both sides. Existing fault tolerance frameworks for LLMs target either training or inference, leaving the optimization potential in the asynchronous execution unexplored for RL. Our key insight is role-based fault isolation so the failure in one machine does not affect the others. We treat trainer, rollout, and other management roles in RL training as distinct distributed sub-tasks. Instead of restarting the entire RL task in ByteRobust, we recover only the failed role and reconnect it to living ones, thereby eliminating the full-restart overhead including rollout replay and initialization delay.

We present RobustRL, the first comprehensive robust system to handle GPU machine errors for RL post-training ETTR (Effective Training Time Ratio) improvement via a \textit{Detect-Restart-Reconnect} paradigm. (1) \textit{Detect}. We implement role-aware monitoring to distinguish actual failures from role-specific behaviors to avoid the false positive and delayed detection. (2) \textit{Restart}. For trainers, we implement a non-disruptive recovery where rollouts persist state and continue trajectory generation, while the trainer is rapidly restored via rollout warm standbys. For rollout, we perform isolated machine replacement without interrupting the RL task. (3) \textit{Reconnect}. We replace static collective communication with dynamic, UCX-based (Unified Communication X) point-to-point communication, enabling immediate weight synchronization between recovered roles. In an RL training task on a 256-GPU cluster with Qwen3-8B-Math workload under 10\% failure injection frequency, RobustRL can achieve an ETTR of over 80\% compared with the 60\% in ByteRobust and achieves 8.4\%-17.4\% faster in end-to-end training time.
\end{abstract}

\section{Introduction}
\label{sec:intro}

Reinforcement Learning (RL) has emerged as a transformative paradigm for post-training LLMs (Large Language Models) \cite{gpt3,gpt4,post-train_survey}, enhancing their reasoning and tool-using capabilities through iterative policy optimization \cite{ppo,grpo,dapo} State-of-the-art models \cite{deepseek-r1,seed-1.6,k1-5,grok4,openai-o1,google-gemini} all leverage RL post-training to achieve superior performance in complex agentic tasks \cite{arxiv25RL-wild}, such as mathematics \cite{AIME}, code generation \cite{deepcoder,swe-bench}, and search \cite{search_r1} \etal.

\begin{figure}[t]
\centering
\includegraphics[width=\linewidth]{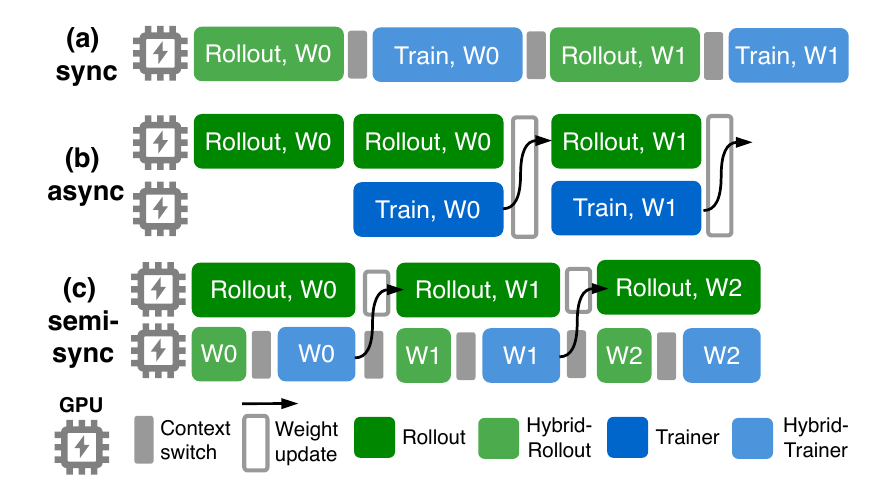}
\vspace{-1.5em}
\caption{Different RL training architectures. The GPU role includes trainer, rollout and hybrid (colocate the both).}	
\vspace{-1.5em}
\label{fig:intro:rl_arch}
\end{figure}

Unlike traditional LLM pre-training \cite{megascale,bytescale,byterobust} or serving \cite{sosp23pagedattention,servegen,kvcacheinthewild,llumnix} workloads, RL post-training comprises two phases: rollout (generation and tool interaction) and training, where they contribute close to 100\% (rollout 80\%) and 70\% (rollout 50\%) post-training time in reasoning tasks and tool learning tasks \cite{arxiv25laminar}. To mitigate the long-tail latency inherent in the rollout phase, RL training frameworks have evolved from synchronous architectures that co-locate rollout and trainers \cite{hybridflow,rlhfuse,openrlhf,deepspeed-chat,realHF} in Figure \ref{fig:intro:rl_arch}(a) to asynchronous designs \cite{areal,streamrl,slime_github,llamarl,asyncflow,arxiv25laminar} in Figure \ref{fig:intro:rl_arch}(b) and (c). These include async mode with dedicated standalone rollouts and trainers, as well as semi-sync mode \cite{seed-1.6,longcat} that combine co-located workers with additional standalone rollouts.

The efficiency of large-scale RL training is therefore critically dependent on the reliability of both the trainer and rollout because they consume the majority of training time \cite{arxiv25laminar}. When scaling to thousands of GPUs, machine failures become the dominant source of task interruption \cite{byterobust,mycroft,osdi25whatif}. However, because rollout and training are tightly interleaved in a single RL task, fault detection and recovery must handle failures stemming from both phases. A naive solution is applying different fault tolerance strategies to the specific roles of the failed components. Unfortunately, they lack the fine-grained mechanisms required for the unique demands of RL. Existing fault tolerance solutions are tailored to pure pre-training \cite{varuna21, easyscale, bamboo, oobleck, parcae, recycle, wagenlander2024tenplex,byterobust} or pure inference workloads \cite{asplos24spotserve,arxiv25expert-service,mooncake}. In current RL systems, the failure of a single critical worker would trigger a RL task restart, discarding training progress and wasting computation resources. Ideally, a fault in a machine hosting one distributed RL role should not disrupt the others. Only the \textit{\textbf{detected}} failed role should \textit{\textbf{restart}} and subsequently \textit{\textbf{reconnect}} with others while the remaining roles continue to run. However, achieving this ideal state presents three significant challenges.

\begin{figure}[t]
\centering
\includegraphics[width=\linewidth]{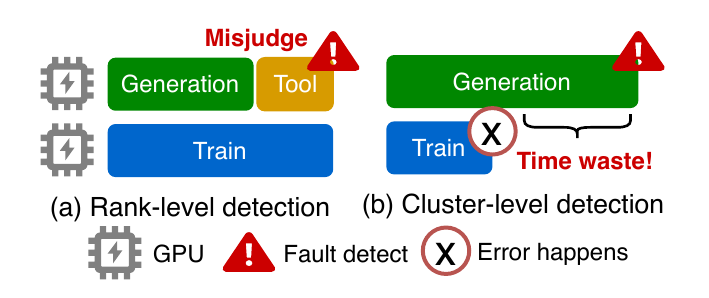}
\vspace{-2.0em}
\caption{Fault detection of pre-train applied in RL. (a) Rank-level leads to false positive. (b) Cluster-level leads to delay.}	
\vspace{-1.5em}
\label{fig:motivation:detect}
\end{figure}

\noindent \textbf{Challenge \#1: Accurate and efficient fault detection.} The rank-level detection for training in ByteRobust does not work for RL. It flags a process as suspect when its behavior deviates from peers or it exhibits zero GPU or network activity \cite{mycroft,osdi25whatif,byterobust,megascale} because each process runs identical forward-backward passes. As shown in Figure \ref{fig:motivation:detect}(a), however, rollouts can idle their GPUs while awaiting external tool responses, causing false positives. Conversely, cluster-level detects the error when all ranks have no GPU activity. It masks these idle periods and introduces significant detection delays as shown in Figure \ref{fig:motivation:detect}. For instance, a trainer network failure might remain undetected for hours until the long-tail rollout phase completes, wasting valuable cluster resources.

\noindent \textbf{Challenge \#2: Fine-grained and efficient restart.} We need to guarantee both machine failures in trainer and rollout would not lead to the RL task restart to preserve the rollout progress. Furthermore, for the trainer, we need rollout to continue the generation during the recovery of the trainer. Besides, the gang-scheduling leads to the long blocking time until the new machine initializes, while the extra warm standby machine leads to the resource waste. An efficient and resource-friendly system is needed for efficient trainer restart. For rollout, we want the recovered rollout to soon get the latest weight for trajectories generation.


\noindent \textbf{Challenge \#3: Dynamic and efficient weight synchronization.} The restart trainer or rollout needs to build reconnection for subsequent weight update as shown in Figure \ref{fig:intro:rl_arch}(b) and (c). Collective communication based on NCCL \cite{nccl} needs to build the fixed communication group before transfer, so it cannot connect with the recovered role in the new machine and does not support fault tolerance. Point-to-point communication like UCX is available for dynamic reconnection, but needs asynchronous and parallel optimization to be efficient.



To address these challenges, we designed \sysname{}, the first comprehensive role-based fault tolerance system for RL post-training in the three RL modes of Figure \ref{fig:intro:rl_arch}. It mitigates the impact of the machine failures for the rollout and trainer. Our contributions are as follows.


\noindent \textbf{1. Role- and phase-aware fault detection.} We introduce a phase-aware and extensible detection strategy for trainer and rollout. Faults in corresponding roles are handled using appropriate robust strategies. (\S\ref{sec:detection}).

\noindent \textbf{2. Role-based and resource-efficient restart.} We propose a decoupled recovery strategy to avoid the RL task restart and preserve the rollout progress. In semi-sync and async mode, for trainer failures, we implement an efficient recovery that dynamically takes rollout as warm standbys, bypassing gang-scheduling delays without requiring extra idle resources. For rollout failures, the recovered rollout would pull the latest weight from other rollouts for trajectory generation immediately. (\S\ref{sec:robust:trainer})


\noindent \textbf{3. Efficient and reliable reconnection.} We implement UCX-based (Unified Communication X) dynamic communication to replace NCCL for weight synchronization. It can achieve efficient weight update through asynchronous point-to-point communication and relay server design. In addition, it can handle the different failure cases during the recovery. (\S\ref{sec:robust:rollout})

\noindent \textbf{4. End-to-end benefits.} In the scenario with frequent failures where a trainer fault is injected every 10\% steps, \sysname{} still maintains an 80\% ETTR on the Qwen3-8B-Math training task with 20\% higher than ByteRobust. The RL training time is 8.4\%-17.4\% faster than ByteRobust. (\S\ref{sec:eval})





%

\section{Background \& Motivation}
\label{sec:bg}

\subsection{RL System for LLM Post-Training}
\label{sec:bg:rlsys}


Post-training using Reinforcement Learning (RL), which evolved from initial human feedback (RLHF) techniques \cite{bai2022antrophic-rlhf,reedGeneralistAgent2022}, is now critical for enhancing the reasoning \cite{grpo,dapo,deepseek-r1} and tool-using capabilities \cite{swe-bench,retool} of LLMs. This process consists of two phases, rollout and training.

\noindent \textbf{Rollout.} The rollout model generates trajectories based on prompts and the results of environment interaction, a process typically completed by an inference engine \cite{vllm,sglang}. For example, mathematical reasoning problems \cite{AIME} can be solved using chain-of-thought (CoT) \cite{wei2022chain}. The rollout model can further achieve tool learning through multi-turn tool invocations. In software engineering \cite{swe-bench} scenarios, trajectories include multi-turn interactions with a code sandbox to fix a bug or implement a specific function. In search scenarios \cite{nq,HotpotQA,search_r1}, trajectories involve multi-turn interactions with a search engine to retrieve target answers.

\noindent \textbf{Training.} The actor updates its policy by trajectory rewards evaluated in the training engine \cite{megatron-lm,fsdp}. The score can be obtained via various reward sources such as rule-based functions \cite{deepseek-r1, dapo} or learned reward models \cite{ouyang2022openai-rlhf, bai2022antrophic-rlhf}. Subsequently, the reference and critic models convert per-trajectory scores and advantages into training experience. Finally, the actor model consumes this experience through a specific policy update strategy \cite{grpo,ppo,dapo,rloo} to complete this iteration.

\noindent \textbf{RL training system.} RL training frameworks like verl \cite{verl} adopt a synchronous mode. After the rollout phase is completed, they switch the inference engine to a training engine by resharding the model weights. However, the multi-turn interactions and decode latency in rollout leads to a significant long-tail latency. Therefore, existing async RL frameworks \cite{streamrl,areal,arxiv25laminar,asyncflow,slime_github} allow the trainer to train within a certain range of offline steps without waiting for all rollout trajectories to complete. It greatly improves training efficiency while maintaining an acceptable loss in precision. Compared to the hybrid deployment of the sync mode, the async paradigm independently deploys the trainer and standalone rollouts, with weight synchronization through network pulling. The semi-sync \cite{longcat,seed1.5} mode replaces the async trainer with a hybrid one, thereby avoiding idle waiting caused by an improper ratio of trainer to standalone rollouts as shown in Figure \ref{fig:intro:rl_arch}.

\subsection{LLM Fault Tolerance System}
\label{sec:bg:fault}


Like pre-training, the RL post-training of large models can last for several weeks or even months \cite{grok4}. During this period, machine failures inevitably occur, leading to a loss of training progress. In RL, a GPU machine where a failure occurs could be executing either a training or an inference task. Prior works, ByteRobust \cite{byterobust} and EaaS \cite{arxiv25expert-service}, have respectively discussed machine failure scenarios during training and inference. In the pre-training process, at the scale of a 100K GPU cluster, machine-induced failures (e.g., GPU, network, CPU, memory, disk) cause over 3,000 interruptions per week on average (30/k-machine/week). For inference services, at the scale of a 4K GPU cluster, 8.8 failures also occur weekly (2.2/k-machines/week), with the main faults manifesting as GPU disconnection and ECC (Error-Correcting Code) errors.

On top of training and inference workload, RL introduces additional stages such as weight synchronization, Ray \cite{ray} management, and tool calls. It further increases the possibility of machine failures. Furthermore, by analyzing issues from open-source RL frameworks with keyword matching, we have found the similar problems caused by the machine error as shown in Table \ref{tab:issue}. Unfortunately, there is no specific robust system for the RL system as we know and the ByteRobust is applied since RL post-training is a kind of training task. Restarting the task when an error happens in ByteRobust is not the best optimization in RL.

\begin{table}[h]
\centering
\footnotesize
\begin{tabular}{l|cccc}
\toprule
Project    & verl \cite{hybridflow}  & Roll \cite{roll} & OpenRLHF \cite{openrlhf_github} & Slime \cite{slime_github} \\ \midrule
Total issues & 1830  & 127  & 719      & 162   \\ \midrule
CUDA error& 189   & 7    & 74       & 14    \\
Timeout  & 152   & 6    & 60       & 9     \\
 OOM      & 117   & 7    & 66       & 12    \\
Job Hang  & 65    & 2    & 20       & 4     \\ \bottomrule
\end{tabular}
\caption{Issues of machine error and job hang in open source RL training framework.}
\label{tab:issue}
\end{table}


For instance, the pre-train robust system ByteRobust \cite{byterobust} achieves warm standby machines to enable fast recovery from a task restart. But it restarts the whole task whatever the error happens, leading to the large progress loss especially in RL training. A better solution is fine-gained restart by role and rollout can be the warm standby for the trainer. We can apply ByteRobust only when the role-based robust strategy in RL cannot handle the case like human-made code or configuration error. For the inference system, they mainly focus on keeping the service running \cite{asplos24spotserve,arxiv25expert-service} or avoid re-prefilling \cite{atc24attentionstore,mooncake} when one inference machine fails. Current RL robust systems follow the inference design \cite{arxiv25laminar,arxiv25rlboost} to achieve the robust rollout because it only needs to extra consider how to reconnect with the trainer in async mode. However, the trainer process has much higher failure frequency compared with inference. The robust trainer is important and the robust rollout should take the trainer failure into consideration.



\begin{figure}[t]
\centering
\includegraphics[width=\linewidth]{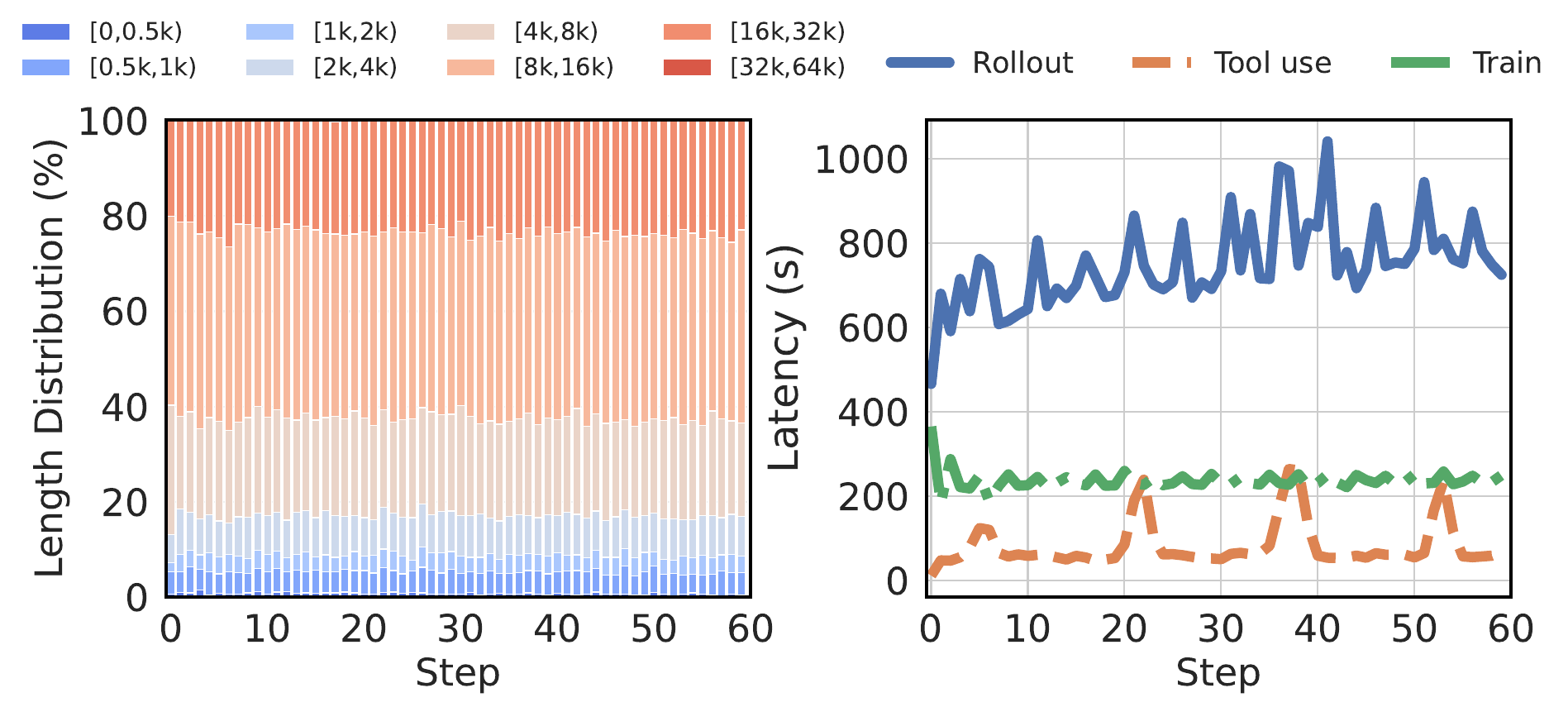}
\vspace{-2.0em}
\caption{Trajectory length distribution (a) and time cost of each step (b) of Search R1 \cite{search_r1} training in HotPotQA dataset \cite{HotpotQA} with Qwen2.5-7B \cite{qwen2.5}.}	
\vspace{-1.0em}
\label{fig:motivation:time}
\end{figure}

\subsection{Opportunity and Challenge}
\label{sec:bg:challenge}

\noindent \textbf{RL robust benefits.} Role-based fault tolerance in RL has two benefits. First, it ensures the failure isolation of different roles. The task can continue to run, which guarantees that the current rollout and agent environment states are maintained. This is important under the long-tail phenomenon, where the rollout phase constitutes the main overhead. In Figure \ref{fig:motivation:time} of search agent training, rollout phase plays the major overhead in one step. Because the response length is long and the tool calling time contributes to the extra overhead. Second, it reduces the recovery time. We only need to restart the trainer instead of the RL task. In the case of async training, the failure of a rollout does not affect the training task, as requests are forwarded to other rollouts. A trainer failure does not affect the rollouts from generating trajectories. To achieve so, we need to handle the following issues.

\noindent \textbf{Per-step checkpoint.} If weights are not saved each step, a weight inconsistency between the recovered trainer and the rollouts will occur when a failure happens. This leads to trajectory deviations and affects the training results. In this situation, it is necessary to roll back to the last checkpoint saved step to retrain and re-rollout. Per-step checkpoint can avoid the re-execution of the time consuming rollout phase.

We believe that a per-step checkpoint is available in the RL scenario based on the following insights. First, in pre-training scenarios, checkpoints are typically saved every tens or hundred steps because a single training step takes only seconds. In contrast, an RL step can last for minutes or even hours \cite{arxiv25RL-wild}, which makes the checkpoint overhead ratio small. Second, in async RL training, the checkpoint process only blocks the trainer for a short time, while rollouts can continue inference during this period. Third, to further reduce the overhead brought by a per-step checkpoint, we apply the ByteCheckpoint \cite{bytecheckpoint}. The blocking time for saving weights is mainly the GPU to memory time within 5 seconds.

\noindent \textbf{Efficient point-to-point communication.} In asynchronous RL training, when trainer or rollout has faults, it needs to re-establish connections after recovery for later weight synchronization shown in Figure \ref{fig:intro:rl_arch}(b) and (c). The existing collective communication protocol NCCL only supports the static communication member so it cannot support robust cases. It gathers the weights of the trainer into rank-0 of sub-group and broadcasts to rank-0 of the inference GPUs. When the number of communication members increases, the transfer overhead would increase linearly. 

We solve this by the UCX-based point-to-point communication. It supports dynamic connection and asynchronous rank-level transmission from the trainer GPU to rollout GPU directly. It solves the traffic bottleneck of single NICs (Network Interface Cards) in NCCL. In addition, the rollout which finishes the weight update can be the relay server for the outdated or recovered rollout pulling. Finally, We further support the fault tolerance case during weight synchronization.


\section{System Overview}
\label{sec:overview}

\begin{figure}[t]
\centering
\includegraphics[width=\linewidth]{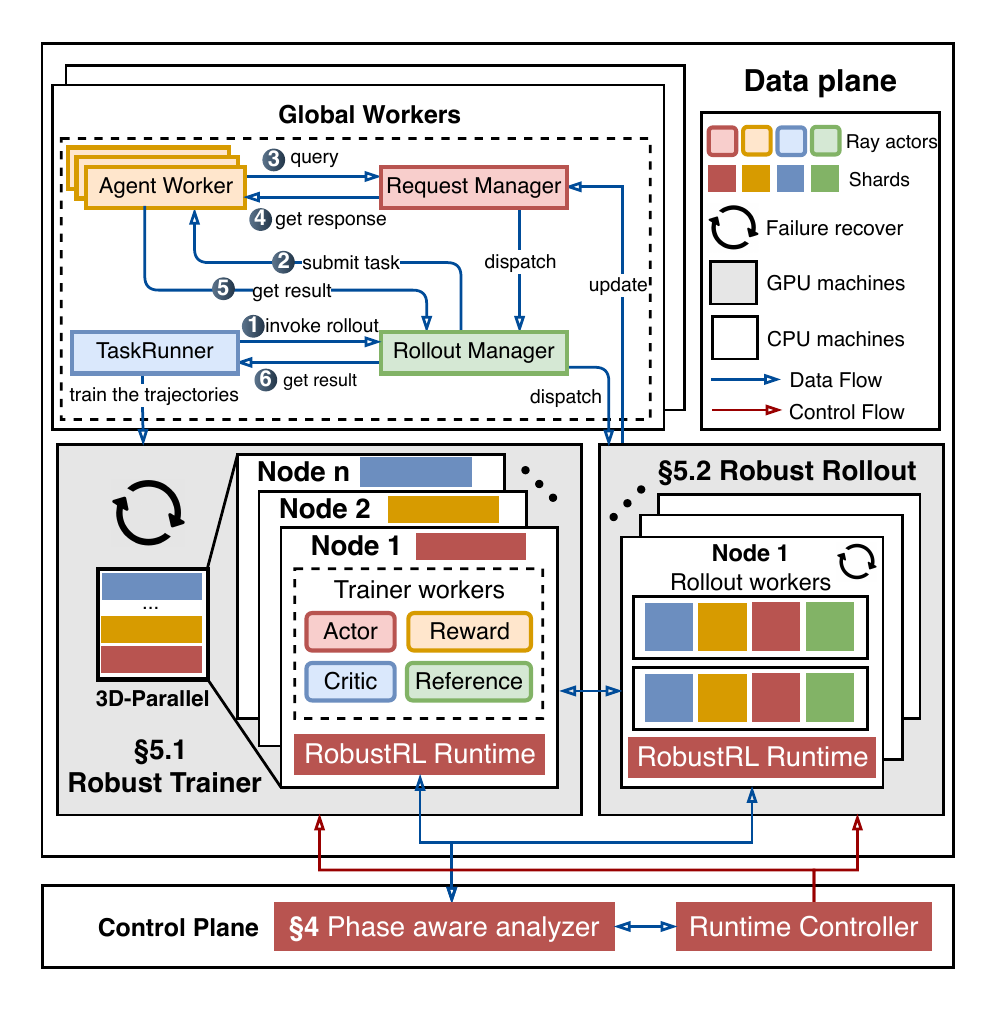}
\vspace{-2.0em}
\caption{System overview of \sysname{}.}	
\label{fig:overview}
\vspace{-1.0em}
\end{figure}

The design goal of \sysname{} is to ensure reliable RL training by using a role-based fault detection mechanism and applying different fault recovery strategies to trainer and rollout. \sysname{} consists of two core components: control plane and data plane. The architecture overview is shown in Figure \ref{fig:overview}.

\noindent \textbf{Control Plane.} It mainly consists of the phase aware analyzer and the runtime controller. The analyzer is responsible for interacting with the RL robust runtime. It tailors fault-detection strategies to each RL role and training stage. The controller, responsible for machine scheduling, receives analysis results from the analyzer and adopts the appropriate fault tolerance and recovery strategy (\S\ref{sec:detection}).

\noindent \textbf{Data Plane.} The GPU machines are the primary failure sources and they all have fault tolerance capabilities (\S\ref{sec:robust}) for machine errors with \sysname{}. All trainers would restart whenever any trainer machine fails, while rollouts would only restart or replace the corresponding faulty machine.

The CPU machine includes management roles, such as the \texttt{AgentWorker} for tool interaction, the \texttt{RolloutManager}, and the \texttt{RequestManager} for managing inference instances and requests. They are placed on CPU machines, and the number of machines is determined by the workload. There can be multiple \texttt{AgentWorkers}, while the other roles default to one instance each. The \texttt{TaskRunner} invokes the \texttt{RolloutManager} to get the trajectories (\ding{192}). Then \texttt{RolloutManager} invokes the \texttt{AgentWorker} (\ding{193}) to call the inference engine and tools to store the result in the \texttt{RequestManager} (\ding{194}). This process can be a multi-turn conversation and the \texttt{AgentWorker} get the trajectories (\ding{195} and \ding{196}) to the trainer (\ding{197}).

A failure in these roles triggers a RL task restart \cite{byterobust}. We use affinity scheduling to ensure that management roles are not scheduled on trainer or rollout machines. This eliminates RL task restarts that would otherwise be triggered by the termination of a management role when a trainer or rollout machine is replaced after a failure.



\section{Role-based Fault Detection}
\label{sec:detection}

\begin{figure}[h]
\centering
\includegraphics[width=\linewidth]{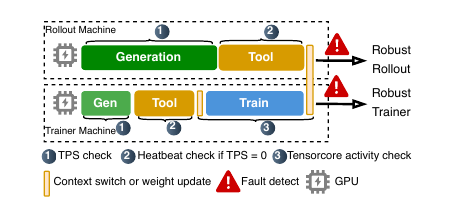}
\vspace{-2.0em}
\caption{Role- and phase-aware fault detection in semi-sync RL training. Row 2 is the hybrid. TPS: Throughput Per Second.}	
\label{fig:phase}
\vspace{-0.5em}
\end{figure}

We illustrate the semi-sync RL workflow in Figure \ref{fig:phase} since both sync and async training are special cases of it. Our goal is to detect which trainer or rollout machine has failed as quickly as possible, so that we can avoid the GPU time waste by restarting or replacing the machine. 


\vspace{-1.0em}

\noindent \textbf{Trainer fault detection.} As introduced in \S\ref{sec:bg:rlsys}, the training phase is divided into the forward-only computation of the reference/critic/reward models and the forward-backward computation of the actor/critic policy update. The detection strategy during this process is consistent with that of pre-training. Failures manifest as zero GPU TensorCore activity for a five-minute window in \sysname{}. When this occurs, we switch to the robust trainer workflow (\S\ref{sec:robust:trainer:workflow}).

Trainer detection is only applied during the training phase on the trainer machines (Figure \ref{fig:phase} \ding{194}). Although phases such as context switching, weight updates, or advantage computation exhibit no GPU activity, their duration is short enough relative to the training phase. Thus, monitoring TensorCore activity is sufficient to detect failures without false positives from short GPU-idle phases. Users can also extend their own detection strategy for this case. For example in large scale RL training, the advantage computation can be over 5 minutes and we can set a large GPU idle threshold. The trainer fault detection method is orthogonal to both pre-training \cite{mycroft,nsdi25minder} and machine diagnostic methods \cite{byterobust}.

\noindent \textbf{Rollout fault detection.} Rollout workloads comprise inference, weight synchronization, and awaiting requests or tool returns. TensorCore activity occurs solely during inference. In the other two stages, the worker exhibits no GPU activity yet remains healthy. We first periodically collect the throughput of each rollout from the \texttt{RolloutManager} (\ding{192}). If the rollout exhibits zero throughput for a specified time interval (60 seconds in our system), we mark it as suspect and trigger heartbeat detection to further verify its status (\ding{193}). No response within the timeout confirms a rollout machine failure and we will handle it with a robust rollout strategy (\S\ref{sec:robust:rollout}).

For the hybrid-rollout of sync and semi-sync mode, this detection method can also be applied. Since the error happens at the trainer machine, we switch to the robust trainer workflow (\S\ref{sec:robust:trainer:workflow}). 



\noindent \textbf{Extensibility to complex faults.} It is worth noting that the role separation architecture proposed in \sysname{} naturally facilitates the integration of advanced fault detection methods. By decoupling the execution contexts, our system becomes inherently compatible with specialized detection strategies designed for complex issues like silent data corruption (SDC) \cite{osdi25trainconfidence,sosp25trainverify,osdi25semanticdetect} and stragglers \cite{osdi25whatif,atc25grayhound} to further harden the system.

\section{Role-based Fault Tolerance}
\label{sec:robust}

\subsection{Robust Trainer}
\label{sec:robust:trainer}

This section illustrates the benefit of robust trainers in different RL architectures (\S\ref{sec:robust:trainer:timeline}), appropriate restart strategy (\S\ref{sec:robust:trainer:workflow}) and reduction of restart overhead by rollout warm standbys (\S\ref{sec:robust:trainer:warm}).

\begin{figure}[t]
\centering
\includegraphics[width=\linewidth]{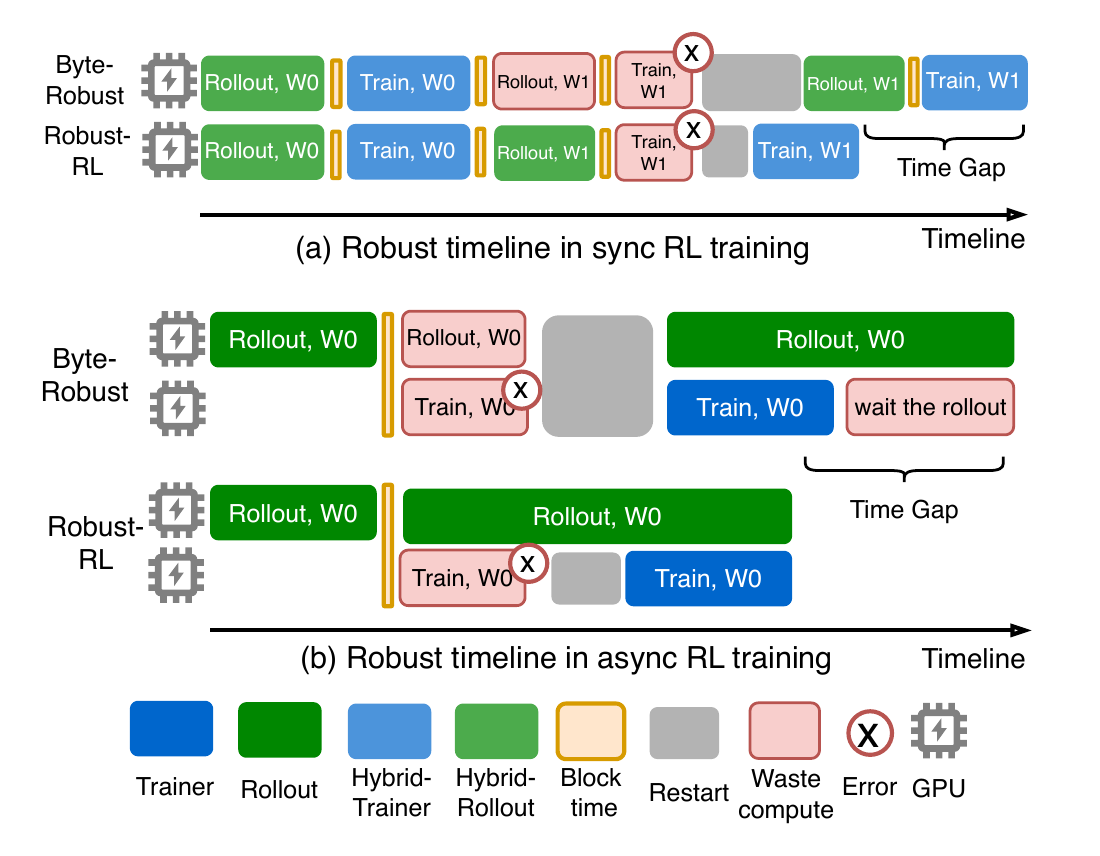}
\vspace{-2.0em}
\caption{Timeline comparison between ByteRobust and \sysname{} in (a) sync and (b) async mode when the trainer machine error happens.}	
\vspace{-1.0em}
\label{fig:trainer:timeline}
\end{figure}

\begin{figure*}[t]
\centering
\includegraphics[width=\linewidth]{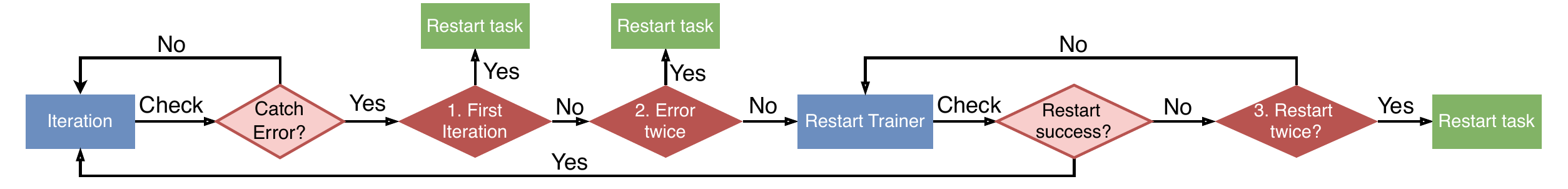}
\caption{Robust trainer workflow. The blue part is the operation in RL training. The red part is the judge for the robust case. The green part is the policy to handle the robust case.}	
\vspace{-1.0em}
\label{fig:robust:workerflow}
\end{figure*}

\subsubsection{Robust Trainer Timeline}
\label{sec:robust:trainer:timeline}

Figure \ref{fig:trainer:timeline} illustrates the difference between the ByteRobust and the \sysname{} in both sync and async training mode. The semi-sync mode is the special case of sync and async mode because it contains both hybrid and standalone rollout. The yellow part in Figure \ref{fig:trainer:timeline} denotes the time that blocks the RL training, which includes weight synchronization, per-step checkpoint and weight reshard between training and inference engine.

\noindent \textbf{Robust trainer benefit.} In sync training mode of Figure \ref{fig:trainer:timeline}(a), ByteRobust restarts the entire task, with the rollout progress of the current iteration lost. \sysname{} only restarts the affected hybrid role and preserves the rollout progress. After recovery, it simply resumes rollout and thus completes the iteration sooner than ByteRobust. In async training mode of Figure \ref{fig:trainer:timeline}(b), the extra benefit is that rollout can continue running during the trainer recovery. Under the same failure case, the rollout in \sysname{} ends earlier, it avoids the trainer to wait for the rollout ends in ByteRobust. In addition, compared with restarting the RL task, we eliminate the overhead for container initialization, ray cluster initialization \etal.

\noindent \textbf{Low blocking overhead.} Due to the long-tail phenomenon of rollout, one step in RL is long shown in Figure \ref{fig:motivation:time}(b), making the blocking time relatively short. The efficiency of rollout weight update has been solved in \S\ref{sec:robust:rollout:ucx}. In addition, we introduce ByteCheckpoint \cite{bytecheckpoint} to improve the checkpoint saving efficiency. The blocking time is only the GPU-to-memory, while the transfer from memory to disk or other persistent storage is executed asynchronously.

\subsubsection{Robust Trainer Workflow}
\label{sec:robust:trainer:workflow}

Robust trainer targets restarts from machine failures and we must avoid an infinite loop of trainer restarts by the human-induced fault. When a human-induced fault occurs, we restart the task and check the code or roll back the code version instead of trainer role restart as soon as possible. This operation "Restart task" is similar to ByteRobust's machine diagnostics and code rollback procedures. We present the robust trainer workflow in Figure \ref{fig:robust:workerflow}.

\noindent \textbf{Iteration.} One iteration includes the step \ding{192} to \ding{197} in Figure \ref{fig:overview}. The RL training is completed through the iteration of multiple steps. Robust trainer wraps the training step function with a \texttt{try-catch} block to capture exceptions during training. The sources of exceptions include both explicit and implicit failures, which are detected by the method in \S\ref{sec:detection}. When the exception happens it turns to the robust phase. 

\noindent \textbf{Trainer restart.} First, \texttt{TaskRunner} terminates all trainer-related processes. Then, it re-executes the trainer's initialization phase and finally reloads the weights from the last saved model checkpoint. Since the model weights are saved each step, the reloaded weights are consistent with the rollout. Information that binds rollouts to the trainer, such as communication addresses for pulling weights, is updated concurrently with the trainer loading the weights.

When we restart to iterate, we skip loading a new batch from the dataset. When the current batch finishes rollout, \texttt{TaskRunner} fetches the trajectories from the \texttt{RequestManager} to the trainer. Otherwise, if we use the sync or semi-sync mode for RL training, we switch the context to the inference engine to rollout the trajectories.

In the following situations, we restart the RL task instead of only the trainers shown in Figure \ref{fig:robust:workerflow}.

\noindent \textbf{1. First-iteration exception detected.} If the model fails on the first step of training or on the first step after resuming training, this indicates an explicit machine or code error. \footnote{Code errors in rollout trigger this process as well.}

\noindent \textbf{2. Repeated exception detected.} If it is the second time the fault has been detected in the current step, we indicate a reproducible fault exists in this step. This situation is also similar to a code error and we restart the task.

\noindent \textbf{3. Repeated restart failure detected.} The restart process itself may fail. For example, if resources are not fully released during the trainer restart or if connection establishment fails due to the restart. In such cases, one restart failure is permitted. If restart failure persists, we also consider it to be a reproducible machine or code error.

\subsubsection{Trainer Warmup by Rollout}
\label{sec:robust:trainer:warm}

\begin{figure}[h]
\centering
\includegraphics[width=\linewidth]{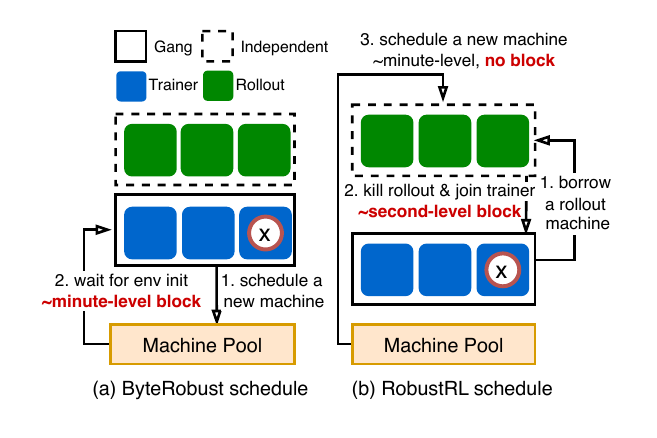}
\vspace{-2.5em}
\caption{Schedule without extra warmup machines when trainer fails. (a) ByteRobust scheduled a new one from the machine pool. (b) \sysname{} borrows a machine from rollout and rollout schedules a new one from the pool.}	
\vspace{-1.0em}
\label{fig:robust:warmup}
\end{figure}

The trainer schedule follows the gang scheduling, meaning the trainer's initialization cannot begin until a new machine is scheduled and the environment is initialized, which is time-consuming as shown in Figure \ref{fig:robust:warmup}(a). ByteRobust prepares the extra warm standby machines to replace faulty ones, thereby saving scheduling and container initialization time. However, using redundant machines leads to potential resource waste, making this strategy limited to large-scale training.

This design does not work for sync-mode training. In semi-sync and async mode, we can use rollouts as warm standbys for the trainer as shown in Figure \ref{fig:robust:warmup}(b). When a failure occurs, one rollout is killed, and it will join the trainer's initialization. Since rollouts use independent scheduling, requests being processed by the rollout on the replaced machine are handled through the robust rollout strategy (\S\ref{sec:robust:rollout}).

The prerequisite for warmup rollout is that rollouts and trainers are in the same data center and rollout and trainer machines are homogeneous. In deployment, similar to ByteRobust, one can reserve a quantity of max(DP, TP, PP, EP, CP) rollouts as homogeneous machines in the same data center, enabling hot-swapping when a failure or suspected fault occurs in any parallel group. Since warmup rollouts have the same environment as the trainer, they can be quickly substituted when a trainer machine fails.


\subsection{Robust Rollout}
\label{sec:robust:rollout}

The design goals for robust rollout are to ensure that after a rollout fails, the recovered rollout can \textbf{1)} re-establish a connection with the trainer. \textbf{2)} Achieve weight high pulling efficiency from the trainer or other relay rollouts. \textbf{3)} Update weights and resume inference as quickly as possible after recovery. To achieve these goals, we have introduced UCX-based weight synchronization (\S\ref{sec:robust:rollout:ucx}) and discuss the situation of the failure handling during rollout recovery (\S\ref{sec:robust:rollout:failure}).


\subsubsection{UCX-based Weight Update}
\label{sec:robust:rollout:ucx}

\begin{figure}[t]
\centering
\includegraphics[width=\linewidth]{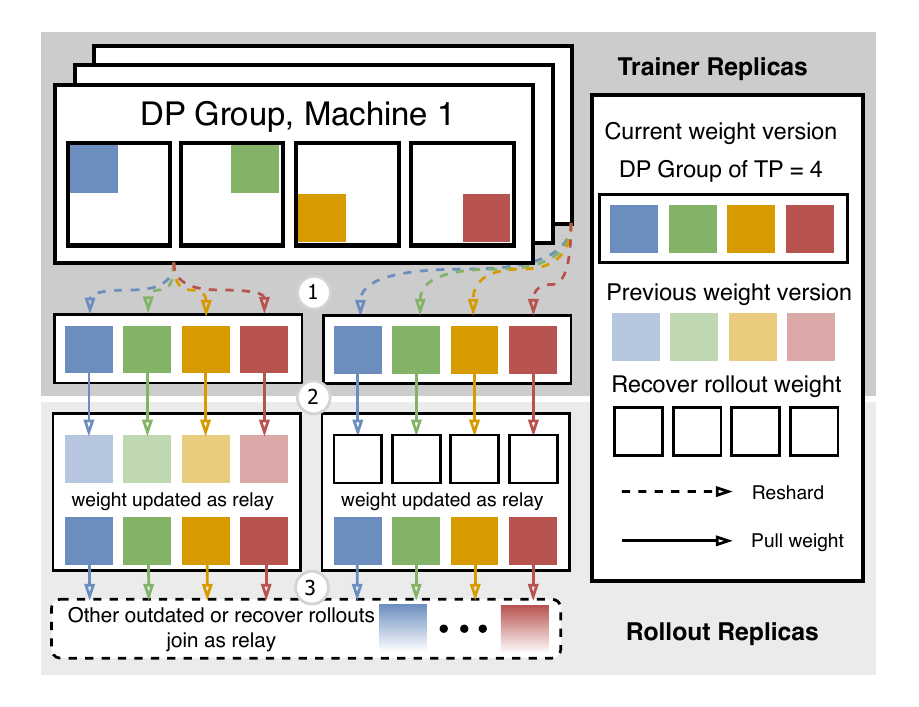}
\vspace{-2.0em}
\caption{UCX-based weight update.}	
\vspace{-1.0em}
\label{fig:rollout:ucx}
\end{figure}

\noindent \textbf{\ding{192} Reshard.} As shown in Figure \ref{fig:rollout:ucx} step \ding{192}, we reshard the model to align with the sharding of the rollout. The implementation is based on verl \cite{verl}, and this process takes under ten seconds.

\noindent \textbf{\ding{193} UCX weight synchronization.} To ensure efficient weight synchronization, we implement point-to-point communication from the trainer's multiple data-parallel (DP) groups to the rollout replicas based on their corresponding ranks shown in Figure \ref{fig:rollout:ucx} \ding{193}. Each transfer initiates by placing the model layer by layer into a buffer for transmission. We ensure the buffer size is large enough to saturate the RDMA bandwidth. Besides, we also prevent OOM error on the GPU by preserving a buffer for transfer. The object pulling weights include both outdated and recovered rollouts awaiting weight updates. We ensure that at any given time, each weight is pulled by only one rollout to prevent network bandwidth contention. 

Compared with current point-to-point communication design \cite{arxiv25laminar}, the entire weight synchronization process is executed on the GPU. However, since torch tensors do not directly support UCX transfer, we convert them to cupy arrays \cite{cupy_learningsys2017}, as cupy supports byte-level transfers. We use DLPack \cite{DLPack} to support a unified tensor memory layout, achieving zero-copy overhead for the conversion from torch tensors to cupy arrays. 

\noindent \textbf{\ding{194} Relay transfer.} A weight updated rollout will act as a relay server, allowing other outdated and recovered rollouts to pull weights from it. Once weight pulling is complete, they join the relay-server set. The pulling process in this stage is the same as the one described in Figure \ref{fig:rollout:ucx} step \ding{194}. If a rollout recovers outside of a weight update stage, it directly pulls weights from a relay server, as shown in step \ding{194}.

The transfer efficiency between trainers and rollouts from different DP groups can vary due to network fluctuations. We make the entire weight synchronization asynchronous. It allows the outdated and recovered rollouts to immediately pull from any rollout having the latest weights. A more extreme asynchronous strategy would be to allow asynchronous transfers at the tensor parallelism (TP) granularity, taking into account that stragglers can exist among different NICs \cite{arxiv25laminar}. However, the async DP transfer for a 235B model on 4$\times$200Gbps NICs is under 10s. This overhead is sufficiently small compared to a single RL step in minutes or even hours level. Therefore, to maintain the simplicity of the weight synchronization design, we did not pursue further optimization. 

In semi-sync RL training, the rollout phase does not begin until all rollouts have finished updating. In this case, the hybrid-rollout in the trainer machine also acts as a relay. In async RL training, the trainer resumes its shard once the first batch of relay rollouts finish their weight update, minimizing training stalls.








\subsubsection{Rollout Failure Cases}
\label{sec:robust:rollout:failure}

\noindent \textbf{Preserve the trajectories.} We save the trajectories of the prompt of each tool iteration in the \texttt{RequestManager} to avoid the rollout progress lost when the rollout machine fails. After each tool iteration, the \texttt{RolloutManager} checks the liveness status of the corresponding rollouts. When a failure occurs on a rollout, the previous results are reassigned to other living rollouts. If it is a rollout machine failure, such as an ECC error, the machine is replaced with a new one. Since the \texttt{AgentWorker} and the sandbox reside outside the rollout machine, the rollout progress can be preserved.

\noindent \textbf{Relay server failure during pulling.} When resuming the weight pull, the rollout will first update its set of target addresses to the living rollouts. If a weight pull fails, it will find another relay server to pull from. The outdated or recovered rollout records their successful pulling progress and pulls the remaining one from other living relay servers.

\noindent \textbf{Trainer failure during pulling.} Failure may happen when the trainer prepares to be pulled or being pulled by rollout. In the first case, rollouts close their established connections with the trainer and wait for its recovery before re-establishing connections. In the case where a pull is in progress, the process is interrupted, and the partially updated weights are cleared. Finally, rollouts wait for the trainer to recover before re-initiating the weight pull.

\section{Implementation}


The implementation of the RL fault tolerance system at the framework level is 8k LoCs of Python for the semi-sync support, fault detection, and robust roles. \sysname{} is built on verl \cite{verl} and its asynchronous RL extension \cite{async-verl}. We extend this foundation with a semi-sync training mode, which retains the trainer's resharding operation to the inference engine. The implementation of the training scheduler that supports the training architectures in Figure \ref{fig:intro:rl_arch}.

\vspace{-1.0em}
\begin{figure}[h]
\centering
\includegraphics[width=\linewidth]{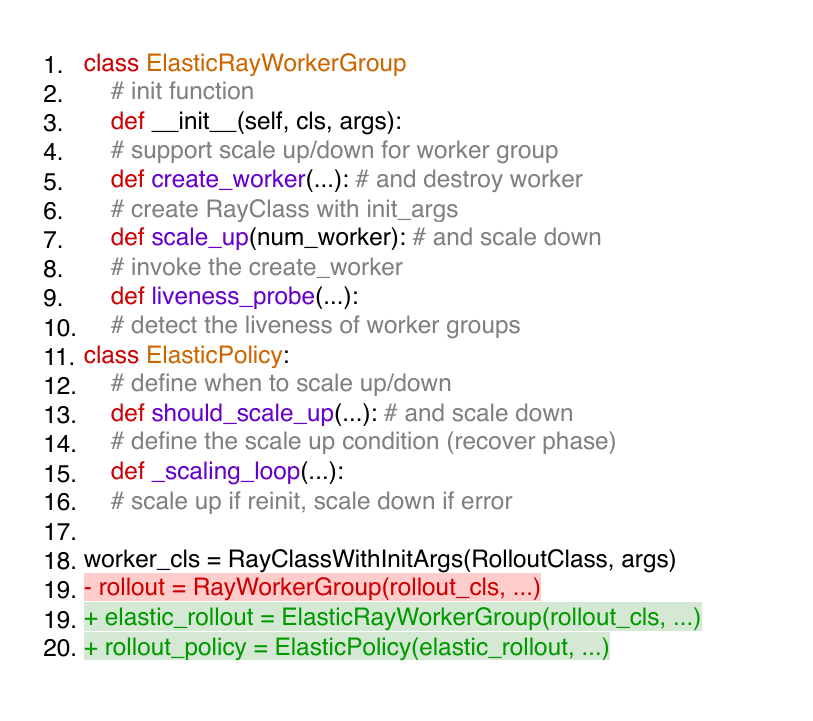}
\vspace{-2.0em}
\caption{The replicated ray worker group and scaling policy for the robust and elastic ray worker in \sysname{}.}	
\vspace{-1.0em}
\label{fig:impl:code}
\end{figure}

\begin{figure*}[h]
\centering
\includegraphics[width=\linewidth]{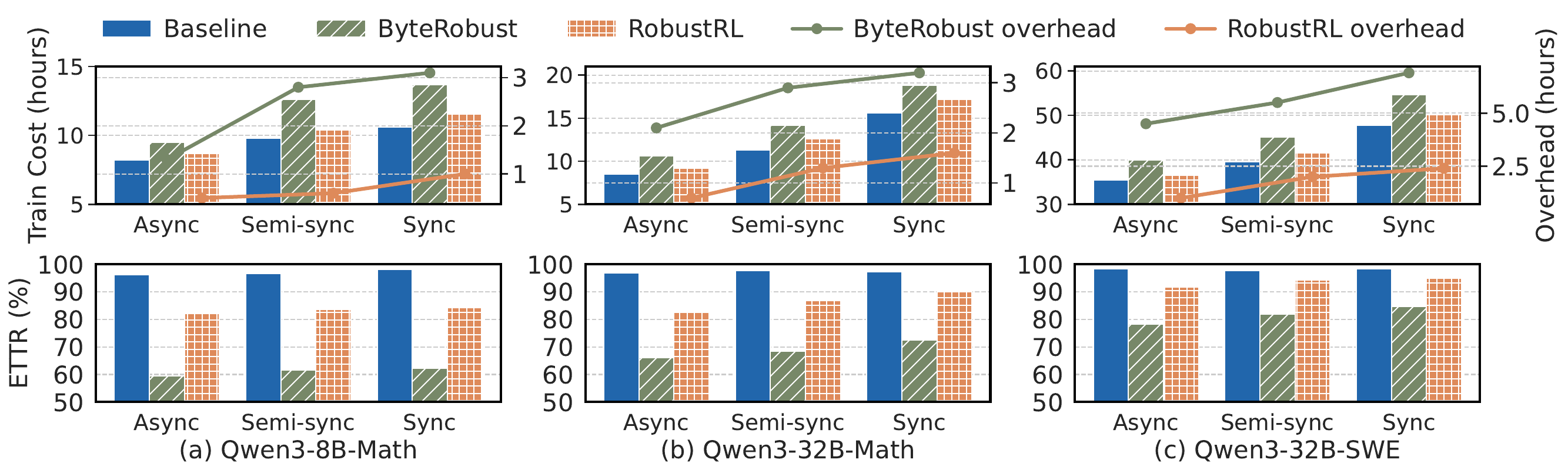}
\vspace{-2.0em}
\caption{End to end evaluation for RL post-training of Qwen3 models in math \cite{dapo} and SWE-bench \cite{swe-bench}. The first row is the job completion time and the second row is the ETTR. The overhead of the first row is the time gap with Baseline.}	
\vspace{-1.5em}
\label{fig:eval:overall}
\end{figure*}

\noindent \textbf{Robust API.} In verl, the initialization of workers for the trainer or rollout roles uses the \texttt{RayWorkerGroup} (RWG) in line 19 of Figure \ref{fig:impl:code}. We further encapsulate this abstraction to handle fault tolerance and recovery scenarios (line 19-20). The \texttt{ElasticRayWorkerGroup} (ERWG) supports expansion and destruction, while a \texttt{ElasticPolicy} determines the timing for these actions. The ERWG supports monitoring the liveness status of the RWG and defines the initialization and destruction functions, \texttt{create\_worker} and \texttt{destroy\_worker}, along with their corresponding pre- and post-processing hooks. The \texttt{scale\_up} is used to determine the number of workers to expand. For example, multiple rollouts fail or are taken to warm up the trainer (\S\ref{sec:robust:trainer:warm}) or scales \cite{arxiv25rlboost}. The policy defines the conditions for elastic scaling with a polling thread to check the liveness of the RWG. For instance, when the platform detects a machine failure during training, it is captured by this policy and scales up the failure worker. This abstraction avoids the need to repeatedly implement creation and destruction logic for different RWGs.



\section{Evaluation}
\label{sec:eval}

Our evaluation tries to answer the following questions.

\noindent \textbf{Q1.} What is the end-to-end training time and ETTR of \sysname{} in scenarios with training failures? (\S\ref{sec:eval:e2e})

\noindent \textbf{Q2.} What are the enhancements in \sysname{}'s fault recovery in terms of restart efficiency and state preservation? (\S\ref{sec:eval:benefit})

\noindent \textbf{Q3.} What is the extra overhead introduced by \sysname{} to implement role-specific fault tolerance? (\S\ref{sec:eval:overhead})

\subsection{Setup}

\noindent \textbf{Testbed.} We deploy \sysname{} on 32 GPU machines with 256 GPUs. Each machine is equipped with eight NVIDIA H20 96GB GPUs. The intra-machine and inter-machine bandwidths are configured with 900 GB/s NVLink and 4$\times$200Gbps NICs, respectively. Each machine has 2TB of memory. Our software versions are CUDA 12.4, PyTorch 2.4.1, and NCCL 2.21.5. We build our system on verl 0.5.x \cite{verl}, using Pytorch FSDP2 \cite{fsdp} and vLLM \cite{vllm} as the training and inference backend. Our checkpointing uses ByteCheckpoint \cite{bytecheckpoint} with HDFS as the storage backend. We selected Qwen3-8B, 32B, and 235B-A22B \cite{qwen3} to evaluate the performance because they are the popular models for RL post-training research in academia and industry. We use the 8B and 32B models to test training performance and provide results from the 235B model to test weight saving and weight synchronization costs. The TP size of rollout is 8 to guarantee each rollout occupies a machine.

\noindent \textbf{Baselines.} The methods we compare are ByteRobust and our \sysname{}. The baseline system is RL training without any injected faults. In ByteRobust and \sysname{}, we inject a fault for the trainer at a random time within every 10\% interval steps by killing all trainers. Such extreme error injection frequency can happen at spot instances or elastic training scenarios. For ByteRobust, we retain only the in-place restart part of a failure, without machine rescheduling. The RL training includes the three architectures in Figure \ref{fig:intro:rl_arch}. Semi-sync switches from rollout mode to train mode when 50\% of the prompts in a batch are completed and the sync mode is 100\%. Async is 0\% with a one-step warmup. For semi-sync and async mode, the number of GPUs for the trainer and rollouts is equal. We do not compare with the current robust rollout RL system \cite{arxiv25laminar,arxiv25rdmap2p} because their codes are not open source. In addition, the robust rollout is a function and the failure or recovery of rollout would not lead to the training bottleneck. For the weight synchronization efficiency, we directly compare with the time of ideal network bandwidth.

\noindent \textbf{Dataset.} For reasoning tasks on math with "DAPO-Math-17K" \cite{dapo}, we use a rule-based reward function to score the trajectories. For the multi-turn tool-learning task, we generate the trajectories by obtaining answers through multi-turn interactions with a sandbox in SWE-bench \cite{swe-bench}. We use GRPO \cite{grpo} for both tasks. RbustRL is independent of the specific RL algorithm and the effectiveness of \sysname{} can generalize to the others. Our global batch size is set to 512, with 64 prompts per batch, and 8 responses generated for each prompt following the previous research \cite{arxiv25laminar}. We set the maximum text length to 65536 and the maximum number of tool interaction turns to 50. We train for 100 steps by default.

\noindent \textbf{Metrics.} Our primary evaluation metrics for fault tolerance are the end-to-end training time and the ETTR. We further evaluate the restart time overhead, additional fault tolerance overhead, and the time saved due to fault tolerance for different methods.

\subsection{End to End Evaluation}
\label{sec:eval:e2e}

We compare the end-to-end training time and ETTR for different training architectures under various GPU workloads on math and SWE post-training tasks on 128 GPUs.

\noindent \textbf{End-to-End Training Time.} As shown in the first row of Figure \ref{fig:eval:overall}, the Qwen3-8B-Math takes the shortest time because the model is small and the post-training task is relatively simple. The SWE task takes the longest time because the introduction of tool interaction significantly lengthens the rollout time. Taking the Qwen3-8B-Math task as an example, the training time varies for different training architectures, where the async mode is faster than the sync one by mitigating long-tail overhead. Therefore, the end-to-end training time continuously increases from the async to the sync mode.

Since we inject a fault every 10 training steps, the end-to-end training times for ByteRobust and \sysname{} are longer than the baseline. The overhead introduced by a fault restart is mainly caused by the restart initialization and the loss of training progress, especially the loss of rollout progress (\S\ref{sec:robust:trainer:timeline}). In contrast, \sysname{} can restart only the trainers, allowing the rollouts to continue executing inference and environment interaction processes. Therefore, the overhead of \sysname{} is smaller than that of ByteRobust. As we shift from async to sync training mode, the impact of the long-tail phenomenon becomes progressively more severe, thus the overhead of ByteRobust gradually increases. In sync mode, the rollout and training phases are executed sequentially as shown in Figure \ref{fig:intro:rl_arch}. By leveraging the role-based fault tolerance strategy, \sysname{} explicitly preserves the rollout results. Consequently, if a failure occurs during the training phase, the system can resume directly from the training step, skipping the redundant rollout phase as shown in Figure \ref{fig:trainer:timeline}(a). This stands in contrast to approaches like ByteRobust, which necessitate restarting the entire RL step. Lastly, the fault tolerance benefits of using \sysname{} on complex tasks like SWE are greater than on math reasoning, because the rollout phase is more time-consuming. The benefit of preserving the rollout progress is more obvious.

In summary, under the extreme condition of a 10\% failure rate, \sysname{} can save 0.8-2.1h, 1.4-1.6h, and 3.5-4.5h in end-to-end training time compared to ByteRobust on the 8B-Math, 32B-Math and 32B-SWE tasks respectively. The restart overhead of \sysname{} for fault tolerance accounts for less than 5\% of the total time, while the ByteRobust introduces 20\% overhead.

\begin{figure}[t]
\centering
\includegraphics[width=\linewidth]{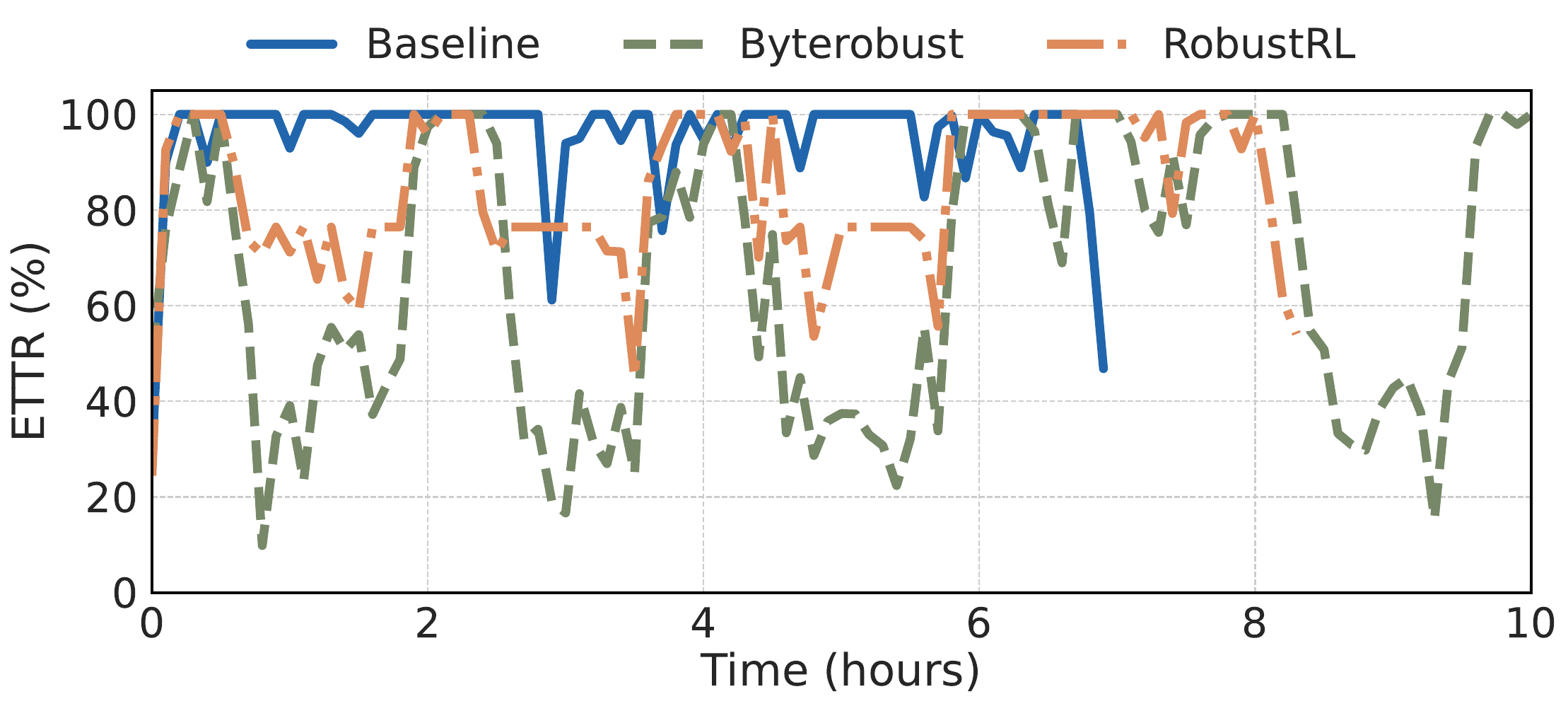}
\vspace{-2.0em}
\caption{Sliding ETTR of the training process Qwen3-8B-Math. The sampling interval is 5 minutes.}	
\vspace{-1.0em}
\label{fig:eval:slide}
\end{figure}

\noindent \textbf{ETTR.} As shown in the second row of Figure \ref{fig:eval:overall}, computation stages such as actor policy updates and rollout are all counted towards the effective training time in RL post-training. In \sysname{}, there is a phase where the trainer restarts for fault tolerance while the rollouts continue running. During this process, the recovery ETTR is calculated with the ratio $ETTR_{ratio}=\frac{\#Rollout}{\#Rollout+\#Trainer}$. Since \sysname{} significantly reduces restart overhead compared to ByteRobust, its ETTR is better than ByteRobust's, and the advantage is more pronounced in simpler tasks like Qwen3-8B-Math. The reason is that the longer training time is, the more negligible the restart time. In addition, although ByteRobust would lose the progress during the restart, the re-execution of rollout is also counted towards ETTR. So the gap of average ETTR between ByteRobust and \sysname{} is not large.

We further show the sliding ETTR for Qwen3-8B-Math in Figure \ref{fig:eval:slide} with semi-sync training mode. We sample the training process every 5 minutes. We take the process of 0.8-2h as an example. Since ByteRobust restarts both the trainer and rollouts when a fault occurs, its ETTR is significantly lower at 20\%. During the same period, \sysname{}'s rollouts continue to execute, so the influence of failure is less. Compared to ByteRobust, \sysname{} improves ETTR by 18-24\% on the Qwen3-8B-Math task.

\begin{figure}[h]
\centering
\includegraphics[width=\linewidth]{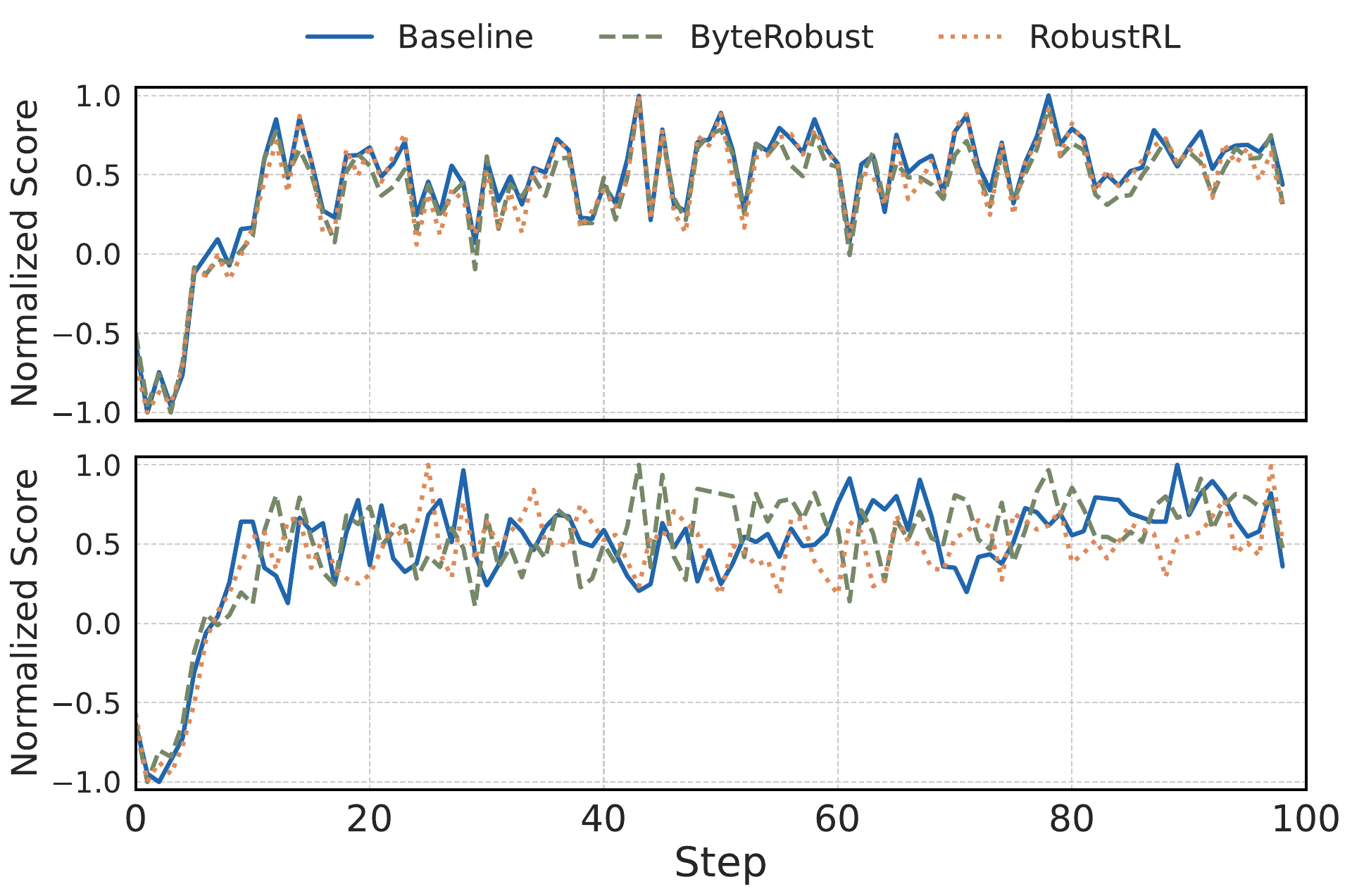}
\vspace{-2.0em}
\caption{Training consistency in sync (row 1) and semi-sync (row 2) mode. The reward of the answer is normalized.}	
\vspace{-1.0em}
\label{fig:eval:consistency}
\end{figure}

\noindent \textbf{Training trend.} We use the training trend of Qwen3-8B-Math with sync and semi-sync to verify the correctness of training in the presence of failures. We execute 100 training steps and obtain curves with similar training trends and the result is shown in Figure \ref{fig:eval:consistency}. In sync training, the three methods has the similar score tendency because we can make sure the training order of the prompt by batch mode instead of the streaming pipeline mode \cite{streamrl}\footnote{Batch mode is waiting for all the prompts in this batch to finish. Streaming mode is adding a new prompt to rollout if one prompt in this batch finishes.}. When the error happens, we can get the prompt and response from \texttt{RequestManager} to continue the training in \sysname{}. However, the result in sync does not get fully alignment because we do not enable the deterministic inference \cite{determisnistic-interence} with FlashAttention \cite{flash-attn} and CUDA atomic operations in our evaluation. In semi-sync situation, the results are still not deterministic and the trend difference is larger because the async schedule mode must use the streaming schedule, which cannot guarantee the order of the prompt. But generally, the three methods in semi-sync training shows the similar tendency.

\begin{figure}[t]
\centering
\includegraphics[width=\linewidth]{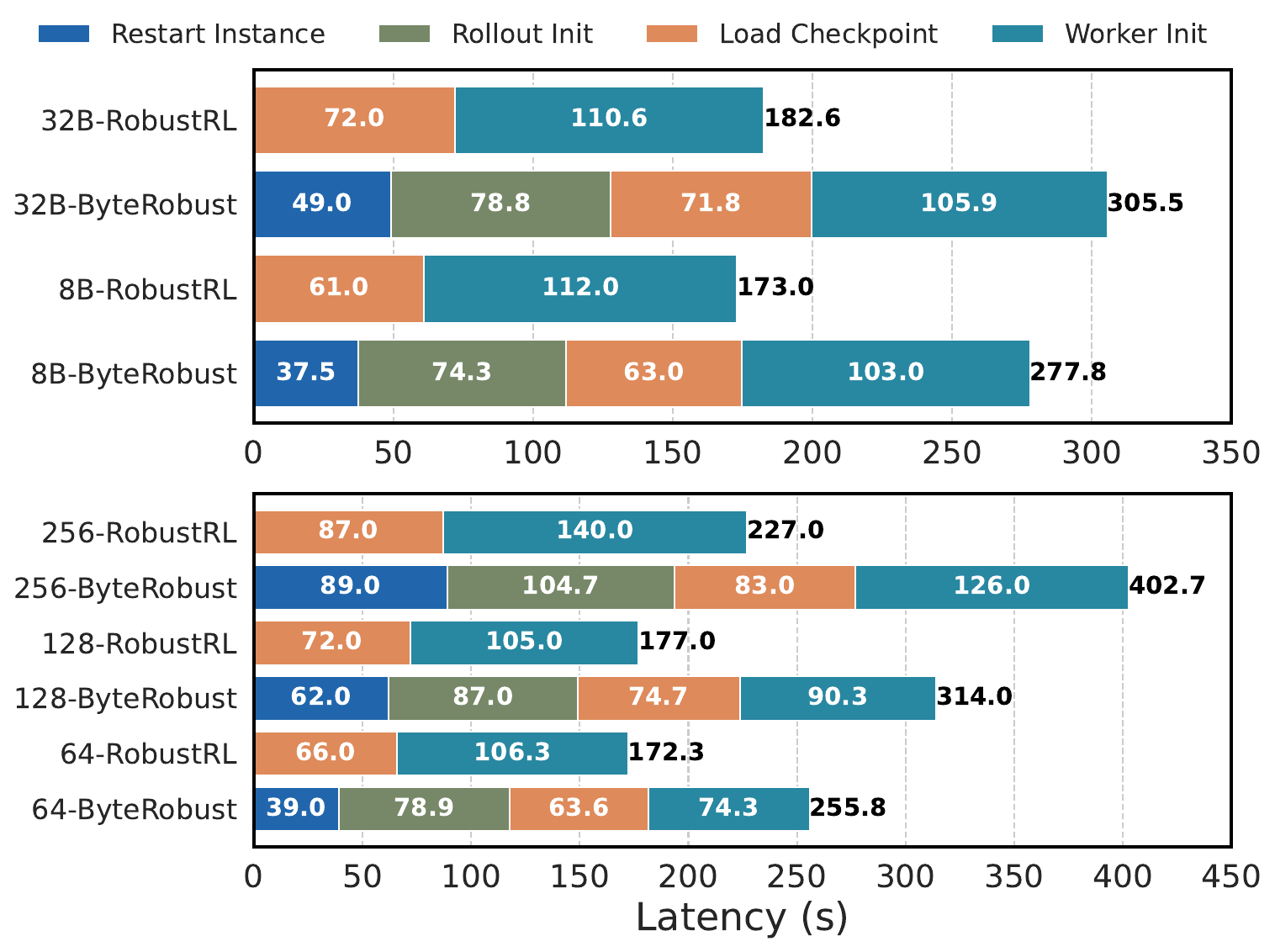}
\vspace{-2.0em}
\caption{Restart cost comparison of ByteRobust and \sysname{}. Row 1 is the result with different model size and row 2 is the result of 8B with different number of GPUs.}	
\vspace{-1.0em}
\label{fig:eval:init}
\end{figure}

\subsection{Robust Benefits Analysis}
\label{sec:eval:benefit}

\noindent \textbf{Trainer restart benefit.} Figure \ref{fig:eval:init} shows a comparison of restart time breakdown between ByteRobust and \sysname{} for different model sizes and numbers of GPUs. We use the semi-sync case as an example because its startup time is the longest, as the trainer machines need both the training and inference engines, in addition to the rollout roles. The main bottlenecks in starting an RL training task consist of four stages: instance restart, rollout initialization, checkpoint loading, and worker initialization. The "restart instance" stage includes container startup, installation of third-party libraries after startup, Kubernetes scheduling \etal. The "worker init" stage is primarily the initialization of the training engine. The "rollout init" stage is the initialization process for the standalone rollout inference engine, not including weight synchronization. The "load checkpoint" is the stage to load from memory to the GPU. When we detect a model anomaly, we asynchronously load the corresponding shards from HDFS to the local memory. The overhead from "rollout init" and "restart instance" phases does not exist in \sysname{}, making it faster than ByteRobust. Benefited by the warmup by rollout mechanism in \S\ref{sec:robust:trainer:warm}, we avoid the trainer's gang scheduling process when the trainer has machine error because we can schedule the rollout machines to replace. 

The second row of Figure \ref{fig:eval:init} shows the results of restart latency of ByteRobust and \sysname{} as the number of GPUs increases, where the duration of all initialization stages increases accordingly. The "worker init" overhead in \sysname{} is larger than in ByteRobust. Compared to ByteRobust's startup, the trainer's restart includes a destruction phase, which introduces additional overhead for destroying network connections. Overall, \sysname{} improves restart efficiency by a factor of $1.5$-$1.7\times$ compared to ByteRobust.

\begin{figure}[t]
\centering
\includegraphics[width=\linewidth]{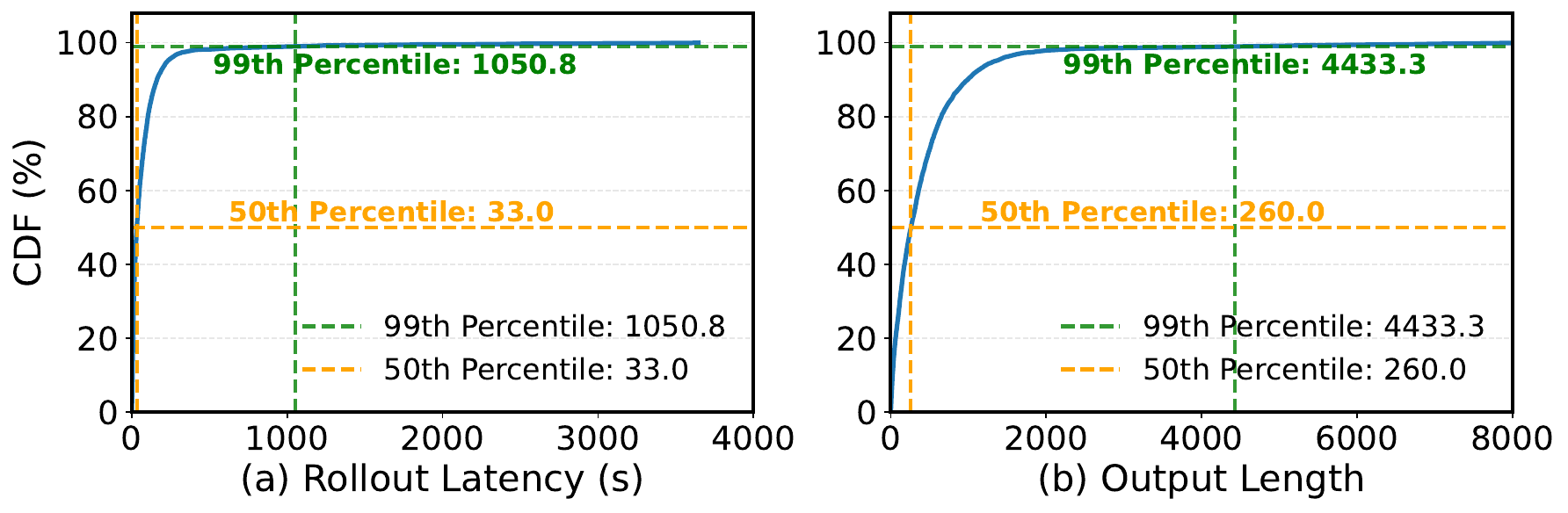}
\caption{Benefit from preserving the rollout progress (a) Rollout cost CDF and (b) Rollout length CDF of all prompts in Qwen3-32B-SWE.}	
\vspace{-1.0em}
\label{fig:eval:benefit}
\end{figure}

\noindent \textbf{Preserving rollout progress benefit.} In addition to the restart benefit, \sysname{} can preserve rollout progress compared to ByteRobust. We recorded the output time distribution and output length distribution for 50k prompts (batch size $\times$ steps = 512 $\times$ 100) on the Qwen3-32B-SWE task shown in Figure \ref{fig:eval:benefit}. The tail latency reaches 1050s, with an output length of over 4k tokens. \sysname{} can avoid the overhead of re-executing this inference. When we extend to larger RL training scenarios involving more conversation turns and more complex tool invocation logic, the benefit can be more significant.

\begin{figure}[t]
\centering
\includegraphics[width=\linewidth]{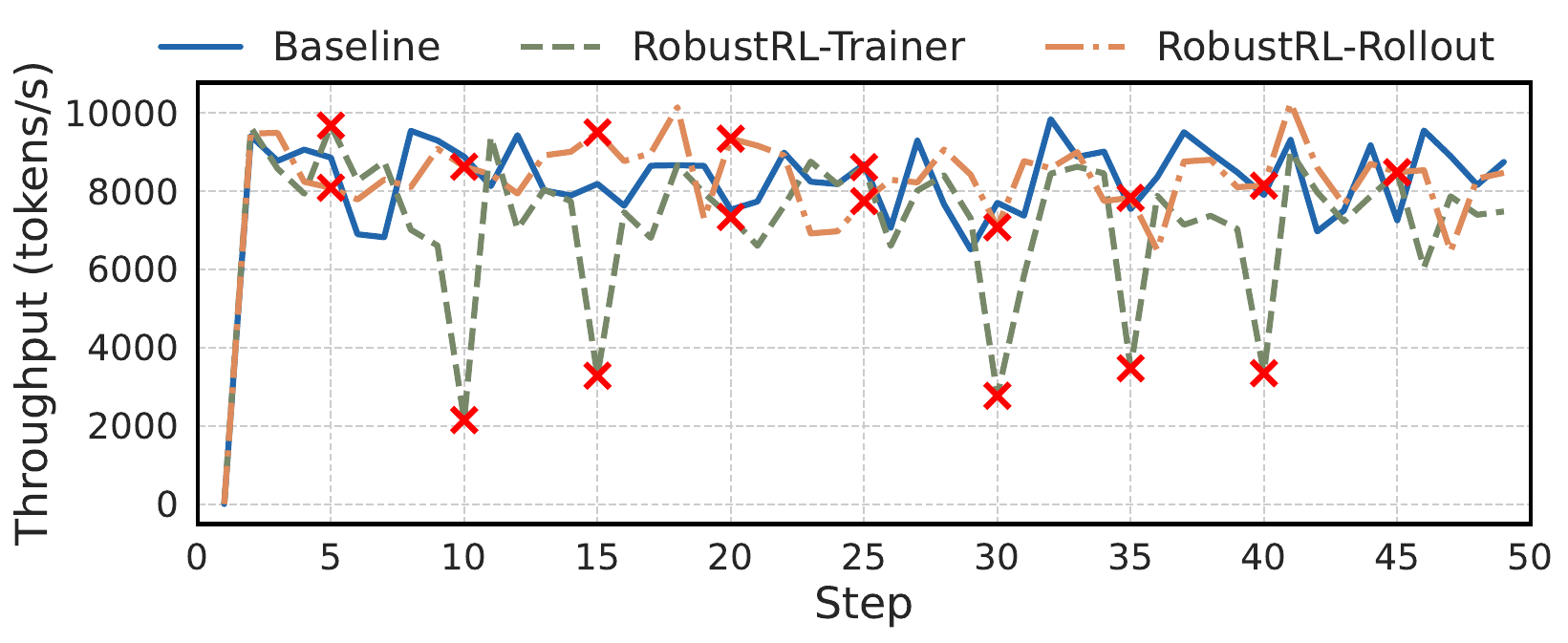}
\vspace{-2.0em}
\caption{Rollout token throughput of Qwen3-8B-Math with 50 steps. We inject the error each five steps.}	
\vspace{-1.5em}
\label{fig:eval:tps}
\end{figure}

\noindent \textbf{Rollout fault tolerance benefit.} A machine failure on a rollout or a machine being borrowed by the trainer does not influence the RL task. There are multiple rollout instances, and in semi-sync mode, the trainer can also provide rollout capabilities. Compared to the over 300s required for a RL task restart shown in Figure \ref{fig:eval:init}, the impact of a rollout restart would not influence the throughput of token decode as shown in Figure \ref{fig:eval:tps}. In the event of a rollout machine failure in 32B model, the time for a single rollout to start up and begin providing service includes machine scheduling (30s), container startup (under 30s), inference engine startup (49s), and weight synchronization (10s), for a total of 119s. 

\noindent \textbf{Throughput under the trainer and rollout failure.} We visualize the rollout throughput of semi-sync training in Figure \ref{fig:eval:tps}. The failure in trainer machines can lead to the throughput decrease because the error can happen when rollout phase finishes. In this phase, the trajectories wait for the consumption by the trainer and continue the next step rollout. The token throughput of rollout failure is not affected because we have multiple rollout replicas.

\noindent \textbf{Detection benefit.} ByteRobust \cite{byterobust} applies the 30 seconds interval for network-related fault detection and 10 seconds for the GPU ones (setup in its Table 1). As discussed in \S\ref{sec:intro}, it would lead to endless restart of the task since the RL task can have idle time waiting for the response from the tools. The \sysname{} can continue the running process. Compared with the cluster-level detection in Figure \ref{fig:motivation:detect}(b), it can save at most about 1000 seconds as shown in Figure \ref{fig:eval:benefit}(a) since we set the hang detection interval for rollout as 60 seconds.

\subsection{Robust Overhead Analysis}
\label{sec:eval:overhead}

The extra overhead for the robust capability includes the UCX weight synchronization and per-step checkpoint.

\begin{figure}[h]
\centering
\includegraphics[width=\linewidth]{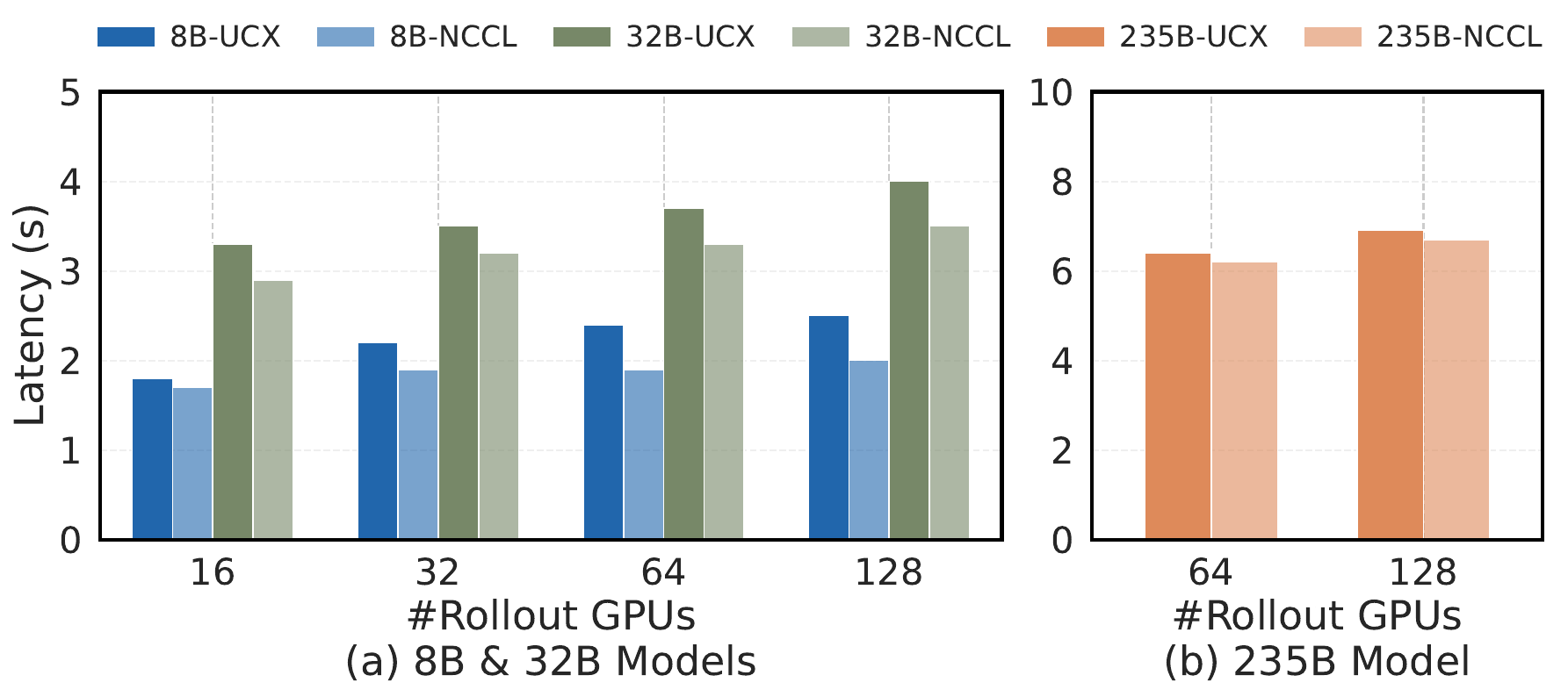}
\vspace{-1.0em}
\caption{Weight synchronization latency with different number of GPUs and model size. The number of GPUs of trainer and rollout is equal. (a) 8B and 32B and (b) 235B.}	

\vspace{-1.0em}
\label{fig:eval:transfer}
\end{figure}

\noindent \textbf{Weight synchronization efficiency.} Figure \ref{fig:eval:transfer} shows the weight synchronization efficiency for an equal number of trainers and rollouts. The 235B model requires more than 64 GPUs for the trainer to start as shown in Figure \ref{fig:eval:transfer}(b). The NCCL communication method first gathers all weight to rank 0 and then broadcasts the weight to all the rollouts. The UCX uses point-to-point communication to the corresponding ranks in rollouts. Since UCX can fully utilize all the bandwidth of the NICs, it can have close performance to the NCCL. In addition, UCX allows for dynamic connections, providing fault tolerance capabilities. For a 235B model with FP16 precision, the model size is 470GB. With the $4\times200$Gbps NICs, the theoretical transfer cost is $\frac{235\times2}{4\times200/8}=4.7$s. Our UCX-based weight synchronization cost is about 6s. The extra overhead can be introduced by the network bandwidth fluctuations and the async bubbles between NICs in one machine.

\begin{figure}[h]
\centering
\includegraphics[width=\linewidth]{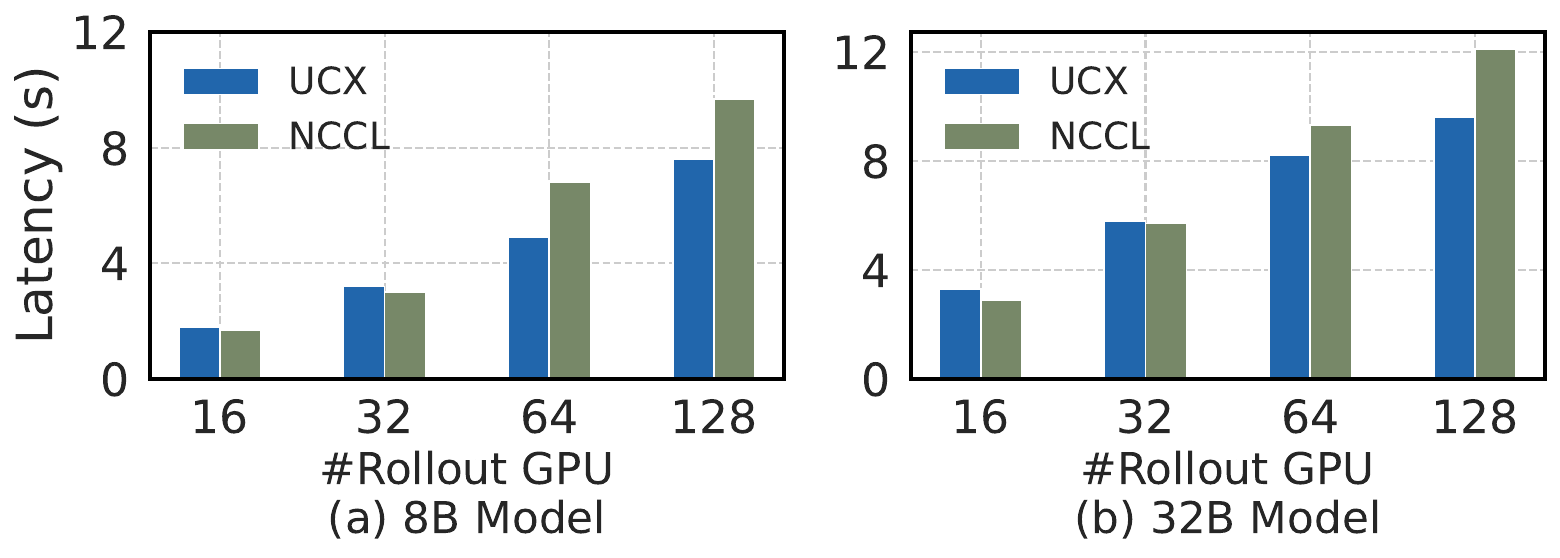}
\vspace{-1.0em}
\caption{Weight synchronization latency with 16 trainer GPUs and different number of rollout GPUs. (a) 8B and (b) 32B models.}	
\vspace{-1.0em}
\label{fig:eval:transfer_rollout}
\end{figure}

We further evaluate the pulling efficiency of our weight synchronization strategy in Figure \ref{fig:eval:transfer_rollout}. Our UCX-based transfer can achieve a linear increase in cost even when the number of rollouts grows exponentially. Benefit by the joining as relay design (\S\ref{sec:robust:rollout:ucx}), the outdated or recovered rollout can pull the weight from the relay server. On the contrary NCCL-based weight synchronization strategy does not support dynamic connection with relay server and all the rollout must pull the weight from trainer. When the number of rollouts is larger than the trainer, its efficiency decreases.

\begin{figure}[h]
\centering
\includegraphics[width=\linewidth]{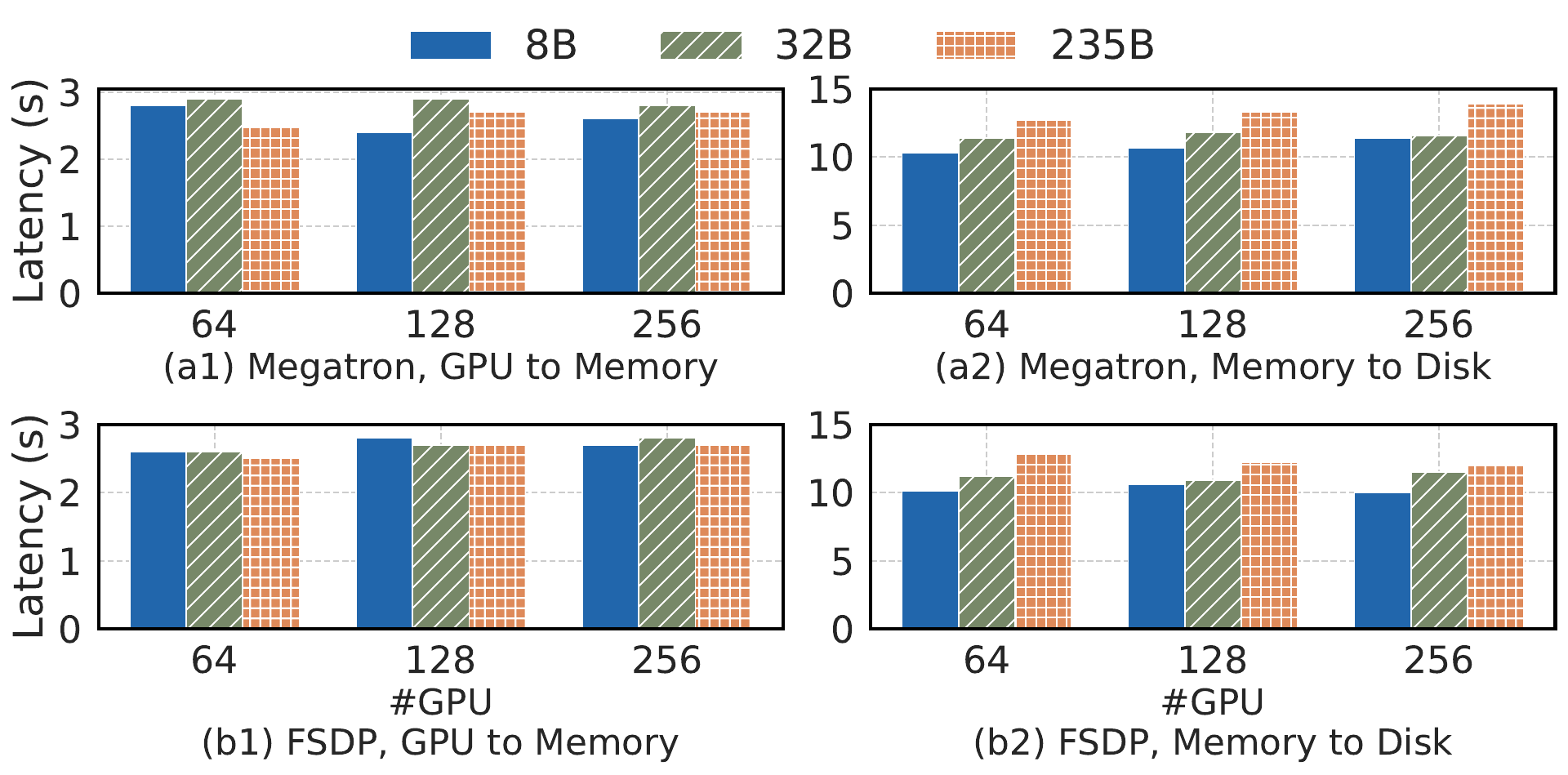}
\vspace{-1.0em}
\caption{Checkpoint latency of ByteCheckpoint. (a) GPU to memory and (b) memory to disk with different model size and number of GPUs.}	
\vspace{-1.0em}
\label{fig:eval:ckpt_latency}
\end{figure}

\noindent \textbf{Checkpoint efficiency.} We evaluate the checkpoint overhead of both FSDP2 and Megatron in Figure \ref{fig:eval:ckpt_latency}. It shows the GPU to memory and memory to disk latency with different model size and number of GPUs for checkpoint. The GPU to memory and memory to disk are independent of the model size and the number of GPUs because the saving process only needs to save the shard corresponding to each rank. The 3s blocking time is 1\% of the minutes or hours required for one step in the RL task. It guarantees that a per-step GPU-to-memory checkpoint would not lead to the OOM during the process of checkpoint from GPU to memory because each step takes over minutes. We have enough time to offload the checkpoint from memory to the disk for about 10s. The disk writing stage is non-blocking, which further minimizes the overhead of checkpoint for the RL training.


\section{Discussion and Limitation}
\noindent \textbf{Limitations of Diagnosis Tools.} Current fault diagnosis tools are primarily designed for training \cite{byterobust,osdi25whatif,wandb,DCGM,mycroft}, lacking specific capabilities for RL post-training scenarios. In RL, the presence of multiple roles and complex data and control dependencies makes root cause localization challenging. For instance, an OOM error in one role might be caused by a memory leak in another co-located role in the same machine. Similarly, a role hanging could be due to abnormal metrics from other roles. More precise RL training system diagnostic tools would help developers pinpoint root causes faster during failures, thereby further improving RL training efficiency.

\noindent \textbf{Role-based Hot Updates.} Adjusting parameters during training is a common requirement. For example, increasing the batch size when GPU memory utilization is found to be low. ByteRobust \cite{byterobust} proposes a hot update mechanism that allows instances to restart and update in-place, avoiding machine rescheduling. Building upon role-based fault tolerance for hot updates, it helps developers achieve faster role-specific parameter updates. \sysname{} reduces restart overhead and prevents the loss of rollout.

\noindent \textbf{Elastic RL Training.} Existing work supports elasticity for the rollout \cite{arxiv25rlboost,streamrl}. Since the rollout performs inference tasks and uses independent scheduling, elastic scaling does not affect task execution. With the support of \sysname{}, the trainer's elasticity in the data-parallel dimension can adopt a similar approach to previous work \cite{gu2023elasticflow,tpds23deepboot,easyscale,zhang2025rubick}, requiring only a trainer restart, thus reducing the overhead of elastic scaling.

\section{Related Work}
\noindent \textbf{Fault tolerance systems for LLM training.} For training systems, their main goal is to locate system failures and recover as soon as possible through fault detection \cite{byterobust,osdi25whatif,mycroft,megascale,nsdi25minder,atc25grayhound}, checkpoint \cite{bytecheckpoint,checkfreq}, and elastic recovery \cite{easyscale,tpds23deepboot} to ensure the correctness of model training and reduce task restart overhead. Some works have further extended to elastic training scenarios, focus on reducing the overhead of scaling due to resource changes. Examples include pipeline redundancy \cite{oobleck} and parallelism adjustment prediction \cite{bamboo} or in spot instance \cite{varuna21}. However, the fault detection strategies for pre-train are not incompatible in RL. Furthermore, they do not consider the recovery of the trainer from partial failures in conjunction with rollouts, nor the design opportunities presented by the asynchronous execution of rollout.

\noindent \textbf{Fault tolerance systems for LLM inference} Existing inference fault tolerance systems primarily operate at the token, and rank levels. The token level refers to utilizing the KV Cache \cite{sosp23pagedattention,atc24attentionstore,mooncake} to avoid re-prefilling requests. The rank level refers to scenarios with AF (attention-FFN) disaggregation \cite{megascale-infer,step3,arxiv25expert-service}, where the failure of a sub-ranks is prevented from causing a full service crash. For example, EaaS considers fault tolerance and recovery after an expert machine fails. However, async RL training scenarios also need to consider the interaction between rollouts and the trainer. Additionally, as an offline training task, we just need to guarantee the prompt can be generated.

\noindent \textbf{RL Systems and Fault Tolerance.} RL involves both training and inference stages. Considering the long-tail phenomenon of rollout, RL training paradigms have shifted from sync \cite{hybridflow,realHF,rlhfuse} to async mode \cite{streamrl,arxiv25laminar,areal,slime_github} to improve training efficiency. Some RL training works have considered the fault tolerance and elasticity of rollouts \cite{arxiv25laminar,arxiv25rlboost}. However, the probability of failure in the rollout phase is much lower than in the training ones because the communication management of training is more complex. \sysname{} further considers the case of trainer fault tolerance, utilizing rollouts as warm standbys and allowing them to continue inference during a trainer failure.

\section{Conclusion}
We have implemented \sysname{}, the first fault tolerance system for RL training that supports all GPU roles against machine failures. Through techniques in \textit{detect-restart-reconnect}, \sysname{} can detect the fault by roles quickly, isolate the failure and minimize the restart overhead with rollout progress preserving. \sysname{} can achieve 80\% ETTR on the 8B-Math training task with 20\% higher than ByteRobust on Qwen3-8B-Math task in extreme robust case.


{\footnotesize \bibliographystyle{acm}
\bibliography{sample}
}

\end{document}